\documentclass[12pt,prd,showpacs,tightenlines,nofootinbib]{revtex4}
\usepackage{bm}
\usepackage{graphics}
\usepackage{rotating}
\usepackage{epsfig}
\begin{document}
\title{\begin{flushright}{\rm\normalsize HU-EP-05/13}\end{flushright}
Masses of light mesons in the relativistic quark model}
\author{D. Ebert}
\affiliation{Institut f\"ur Physik, Humboldt--Universit\"at zu Berlin,
Newtonstr. 15, D-12489  Berlin, Germany}
\author{R. N. Faustov}
\author{V. O. Galkin}
\affiliation{Institut f\"ur Physik, Humboldt--Universit\"at zu Berlin,
Newtonstr. 15, D-12489 Berlin, Germany}
\affiliation{Dorodnicyn Computing Centre, Russian Academy of Sciences,
  Vavilov Str. 40, 119991 Moscow, Russia}

\begin{abstract}
The masses of the $S$-wave mesons consisting of the light ($u,d,s$)
quarks are calculated within the constituent quark model. The
relativistic Schr\"odinger-like equation with a confining potential
 is numerically solved for the
complete relativistic $q\bar q$ potential including both
spin-independent and spin-dependent terms. The obtained masses of the
ground-state $\pi$, $\rho$, $K$, $K^*$ and $\phi$ mesons and their
first radial excitations are in reasonably good overall agreement with
experimental data. 
\end{abstract}


\maketitle

The theoretical description of the properties of light mesons such
as  $\pi$, $\rho$, $K$, $K^*$ and $\phi$ within the constituent quark
model presents additional difficulties compared to heavy-light mesons
and heavy quarkonia. In fact, due to the highly relativistic dynamics
of light quarks, the $v/c$ and $1/m_q$ expansions are completely
inapplicable in 
the case of light mesons and the QCD coupling constant $\alpha_s$ at
the related scale $\mu$ is rather
large. Moreover, the behaviour of $\alpha_s(\mu^2)$ is unknown in
the infrared region and is thus model dependent (exhibiting, e.g.,
freezing behaviour, etc.). The pseudoscalar mesons $\pi$ and $K$
produce a special problem, 
since their small masses originate from their Goldstone nature.

Different attempts to study light mesons on the basis of the
relativized quark model \cite{gi}, the
Dyson-Schwinger and Bethe-Salpeter equations \cite{mr,k}, chiral quark
models with spontaneous symmetry breaking (e.g. the Nambu-Jona-Lasinio
model) \cite{erv} and lattice QCD \cite{ak} were undertaken. 
Here we consider the possibility of describing light mesons on the
basis of the three-dimensional relativistic wave equation with a
confining potential. While
describing the properties of heavy-light 
mesons \cite{egf}, we treated the light quarks in a completely relativistic
way. Now we apply this approach to calculating the  masses of light
mesons in the framework of the previously developed relativistic
quark model.

  In this approach a meson is described by the wave
function of the bound quark-antiquark state, which satisfies the
quasipotential equation  of the Schr\"odinger type~\cite{efg}
\begin{equation}
\label{quas}
{\left(\frac{b^2(M)}{2\mu_{R}}-\frac{{\bf
p}^2}{2\mu_{R}}\right)\Psi_{M}({\bf p})} =\int\frac{d^3 q}{(2\pi)^3}
 V({\bf p,q};M)\Psi_{M}({\bf q}),
\end{equation}
where the relativistic reduced mass is
\begin{equation}
\mu_{R}=\frac{E_1E_2}{E_1+E_2}=\frac{M^4-(m^2_1-m^2_2)^2}{4M^3},
\end{equation}
and $E_1$, $E_2$ are given by
\begin{equation}
\label{ee}
E_1=\frac{M^2-m_2^2+m_1^2}{2M}, \quad E_2=\frac{M^2-m_1^2+m_2^2}{2M}.
\end{equation}
Here $M=E_1+E_2$ is the meson mass, $m_{1,2}$ are the quark masses,
and ${\bf p}$ is their relative momentum.  
In the center-of-mass system the relative momentum squared on mass shell 
reads
\begin{equation}
{b^2(M) }
=\frac{[M^2-(m_1+m_2)^2][M^2-(m_1-m_2)^2]}{4M^2}.
\end{equation}

The kernel 
$V({\bf p,q};M)$ in Eq.~(\ref{quas}) is the quasipotential operator of
the quark-antiquark interaction. It is constructed with the help of the
off-mass-shell scattering amplitude, projected onto the positive
energy states. 
Constructing the quasipotential of the quark-antiquark interaction, 
we have assumed that the effective
interaction is the sum of the usual one-gluon exchange term with the mixture
of long-range vector and scalar linear confining potentials, where
the vector confining potential
contains the Pauli interaction. The quasipotential is then defined by
  \begin{equation}
\label{qpot}
V({\bf p,q};M)=\bar{u}_1(p)\bar{u}_2(-p){\mathcal V}({\bf p}, {\bf
q};M)u_1(q)u_2(-q),
\end{equation}
with
$${\mathcal V}({\bf p},{\bf q};M)=\frac{4}{3}\alpha_sD_{ \mu\nu}({\bf
k})\gamma_1^{\mu}\gamma_2^{\nu}
+V^V_{\rm conf}({\bf k})\Gamma_1^{\mu}
\Gamma_{2;\mu}+V^S_{\rm conf}({\bf k}),$$
where $\alpha_S$ is the QCD coupling constant, $D_{\mu\nu}$ is the
gluon propagator in the Coulomb gauge
\begin{equation}
D^{00}({\bf k})=-\frac{4\pi}{{\bf k}^2}, \quad D^{ij}({\bf k})=
-\frac{4\pi}{k^2}\left(\delta^{ij}-\frac{k^ik^j}{{\bf k}^2}\right),
\quad D^{0i}=D^{i0}=0,
\end{equation}
and ${\bf k=p-q}$; $\gamma_{\mu}$ and $u(p)$ are 
the Dirac matrices and spinors
\begin{equation}
\label{spinor}
u^\lambda({p})=\sqrt{\frac{\epsilon(p)+m}{2\epsilon(p)}}
\left(
\begin{array}{c}1\cr {\displaystyle\frac{\bm{\sigma}
      {\bf  p}}{\epsilon(p)+m}}
\end{array}\right)\chi^\lambda,
\end{equation}
with $\epsilon(p)=\sqrt{p^2+m^2}$.
The effective long-range vector vertex is
given by
\begin{equation}
\label{kappa}
\Gamma_{\mu}({\bf k})=\gamma_{\mu}+
\frac{i\kappa}{2m}\sigma_{\mu\nu}k^{\nu},
\end{equation}
where $\kappa$ is the Pauli interaction constant characterizing the
anomalous chromomagnetic moment of quarks. Vector and
scalar confining potentials in the nonrelativistic limit reduce to
\begin{eqnarray}
\label{vlin}
V^V_{\rm conf}(r)&=&(1-\varepsilon)(Ar+B),\nonumber\\ 
V^S_{\rm conf}(r)& =&\varepsilon (Ar+B),
\end{eqnarray}
reproducing 
\begin{equation}
\label{nr}
V_{\rm conf}(r)=V^S_{\rm conf}(r)+V^V_{\rm conf}(r)=Ar+B,
\end{equation}
where $\varepsilon$ is the mixing coefficient. 

The light constituent quark masses $m_u=m_d=0.33$ GeV, $m_s=0.5$ GeV and
the parameters of the linear potential $A=0.18$ GeV$^2$ and $B=-0.3$ GeV
have the usual values of quark models.  The value of the mixing
coefficient of vector and scalar confining potentials $\varepsilon=-1$
has been determined from the consideration of charmonium radiative
decays \cite{efg}. 
Finally, the universal Pauli interaction constant $\kappa=-1$ has been
fixed from the analysis of the fine splitting of heavy quarkonia ${
}^3P_J$- states \cite{efg}. In the literature it is widely discussed
the 't~Hooft-like interaction between quarks induced by instantons \cite{dk}.
This interaction can be partly described by introducing the quark
anomalous chromomagnetic moment having an approximate value
$\kappa=-0.744$ (Diakonov \cite{dk}). This value is of the same
sign and order of magnitude 
as the Pauli constant $\kappa=-1$ in our model. Thus the Pauli term
incorporates at least part of the instanton contribution to the $q\bar q$
interaction.\footnote{As is well-known, the instanton-induced 't~Hooft
  interaction term breaks the axial $U_A(1)$-symmetry, the violation
of  which is needed for  describing the $\eta-\eta'$ mass splitting. We
do not consider this issue here.}

The quasipotential (\ref{qpot}) can  be used for arbitrary quark
masses.  The substitution 
of the Dirac spinors (\ref{spinor}) into (\ref{qpot}) results in an extremely
nonlocal potential in the configuration space. Clearly, it is very hard to 
deal with such potentials without any additional transformations.
 In oder to simplify the relativistic $q\bar q$ potential, we make the
following replacement in the Dirac spinors:
\begin{equation}
  \label{eq:sub}
  \epsilon_{1,2}(p)=\sqrt{m_{1,2}^2+{\bf p}^2} \to E_{1,2}
\end{equation}
(see the discussion of this point in \cite{egf}).  This substitution
makes the Fourier transformation of the potential (\ref{qpot}) local.
We also limit our consideration only to the $S$-wave
states, which further simplifies our analysis, since all terms 
proportional to ${\bf L}^2$ vanish as well as the spin-orbit
ones. Thus we neglect the mixing of states with different values of
$L$.  Calculating the potential, we keep only  operators quadratic
in the momentum acting on $V_{\rm Coul}$, $V^{V,S}_{\rm conf}$  and
replace ${\bf p}^2\to E_{1,2}^2-m_{1,2}^2$ in higher order operators
in accord  with Eq.~(\ref{eq:sub}) preserving the symmetry under the
$(1\leftrightarrow 2)$ exchange.  

The substitution (\ref{eq:sub})
works well for the confining part of the potential. However, it leads to 
a fictitious singularity $\delta^3({\bf r})$  at the origin arising from the  
one-gluon exchange part ($\Delta V_{\rm
  Coul}(r)$), which is absent in the initial potential.
Note that this singularity is not important if it is treated
perturbatively. Since we are not using the  expansion in $v/c$ and
are solving the quasipotential equation with the 
complete relativistic potential, an additional analysis is
required. Such singular contributions emerge from the following  terms  
\begin{eqnarray}
  \label{eq:st}
 && \frac{{\bf k}^2}{[\epsilon_i(q)(\epsilon_i(q)+m_i)
\epsilon_i(p)(\epsilon_i(p)+m_i)]^{1/2}}V_{\rm Coul}({\bf k}^2) ,\cr
&&\frac{{\bf k}^2}{[\epsilon_1(q)\epsilon_1(p)
\epsilon_2(q)\epsilon_2(p)]^{1/2}}V_{\rm Coul}({\bf k}^2),
\end{eqnarray}
if we simply replace $\epsilon_{1,2}\to E_{1,2}$. However, the Fourier
transforms of expressions (\ref{eq:st}) are less singular at $r\to
0$. To avoid such fictitious singularities we note that if the binding effects 
are taken into account, it is necessary to replace $\epsilon_{1,2}
\to E_{1,2}-\eta_{1,2}V$, where $V$ is the quark interaction potential
and $\eta_{1,2}=m_{2,1}/(m_1+m_2)$. At small
distances  $r\to 0$, the Coulomb singularity in $V$ dominates
and gives the correct asymptotic behaviour. Therefore, we replace
$\epsilon_{1,2} \to E_{1,2}-\eta_{1,2}V_{\rm Coul}$  in  the Fourier
transforms of terms (\ref{eq:st}) (cf. \cite{bs}). We used
the similar regularization of singularities in the analysis of
heavy-light meson spectra \cite{egf}. Finally, we ignore the annihilation
terms in the quark potential since they contribute only in the
isoscalar channels and are suppressed in the $s\bar s$ vector channel.

The resulting $q\bar q$ potential then reads
\begin{equation}
  \label{eq:v}
  V(r)= V_{\rm SI}(r)+ V_{\rm SD}(r),
\end{equation}
where the spin-independent potential for $S$-states (${\bf
  L}^2=0$) has the form 
\begin{eqnarray}
  \label{eq:vsi}
  V_{\rm SI}(r)&=&V_{\rm Coul}(r)+V_{\rm conf}(r)+
\frac{(E_1^2-m_1^2+E_2^2-m_2^2)^2}{4(E_1+m_1)(E_2+m_2)}\Biggl\{
\frac1{E_1E_2}V_{\rm Coul}(r)\cr
&& +\frac1{m_1m_2}\Biggl(1+(1+\kappa)\Biggl[(1+\kappa)\frac{(E_1+m_1)(E_2+m_2)}
{E_1E_2}\cr
&&-\left(\frac{E_1+m_1}{E_1}+\frac{E_1+m_2}{E_2}\right)\Biggr]\Biggr)
V^V_{\rm conf}(r)
+\frac1{m_1m_2}V^S_{\rm conf}(r)\Biggr\}\cr
&&+\frac14\left(\frac1{E_1(E_1+m_1)}\Delta
\tilde V^{(1)}_{\rm Coul}(r)+\frac1{E_2(E_2+m_2)}\Delta
\tilde V^{(2)}_{\rm Coul}(r)\right)\cr
&&-\frac14\left[\frac1{m_1(E_1+m_1)}+\frac1{m_2(E_2+m_2)}-(1+\kappa)
\left(\frac1{E_1m_1}+\frac1{E_2m_2}\right)\right]\Delta V^V_{\rm
conf}(r)\cr
&&+\frac{(E_1^2-m_1^2+E_2^2-m_2^2)}{8m_1m_2(E_1+m_1)(E_2+m_2)} 
\Delta V^S_{\rm conf}(r), 
\end{eqnarray}
and the spin-dependent potential is given by
\begin{eqnarray}
  \label{eq:vsd}
   V_{\rm SD}(r)&=&\frac2{3E_1E_2}\Biggl[\Delta \bar V_{\rm Coul}(r)
+\left(\frac{E_1-m_1}{2m_1}-(1+\kappa)\frac{E_1+m_1}{2m_1}\right)\cr
&&\qquad\quad\times
\left(\frac{E_2-m_2}{2m_2}-(1+\kappa)\frac{E_2+m_2}{2m_2}\right)
\Delta V^V_{\rm conf}(r)\Biggr]{\bf S}_1{\bf S}_2,
\end{eqnarray}
with
\begin{eqnarray}
  \label{eq:tv}
V_{\rm Coul}(r)&=&-\frac43\frac{\alpha_s}{r},\cr
\tilde V^{(i)}_{\rm Coul}(r)&=&V_{\rm Coul}(r)\frac1{\displaystyle\left(1+
\eta_i\frac43\frac{\alpha_s}{E_i}\frac1{r}\right)\left(1+
\eta_i\frac43\frac{\alpha_s}{E_i+m_i}\frac1{r}\right)},\qquad (i=1,2),\cr
  \bar V_{\rm Coul}(r)&=&V_{\rm Coul}(r)\frac1{\displaystyle\left(1+
\eta_1\frac43\frac{\alpha_s}{E_1}\frac1{r}\right)\left(1+
\eta_2\frac43\frac{\alpha_s}{E_2}\frac1{r}\right)}, 
\qquad \eta_{1,2}=\frac{m_{2,1}}{m_1+m_2}.
\end{eqnarray}
Here we put  $\alpha_s\equiv\alpha_s(\mu_{12}^2)$ with $\mu_{12}=2m_1
m_2/(m_1+m_2)$. We adopt for $\alpha_s(\mu^2)$ the
simplest model with freezing \cite{bvb}, namely
\begin{equation}
  \label{eq:alpha}
  \alpha_s(\mu^2)=\frac{4\pi}{\displaystyle\beta_0
\ln\frac{\mu^2+M_B^2}{\Lambda^2}}, \qquad \beta_0=11-\frac23n_f,
\end{equation}
where the background mass is $M_B=2.24\sqrt\sigma=0.95$~GeV \cite{bvb}, and
$\Lambda=413$~MeV was fixed from fitting the $\rho$ mass. We put the
number of flavours $n_f=2$ for $\pi$, 
$\rho$, $K$, $K^*$ and $n_f=3$ for $\phi$. As a result we obtain
$\alpha_s(\mu_{ud}^2)=0.730$, $\alpha_s(\mu_{us}^2)=0.711$ and
$\alpha_s(\mu_{ss}^2)=0.731$.   

\begin{figure}
\centering
  \includegraphics[width=7.7cm]{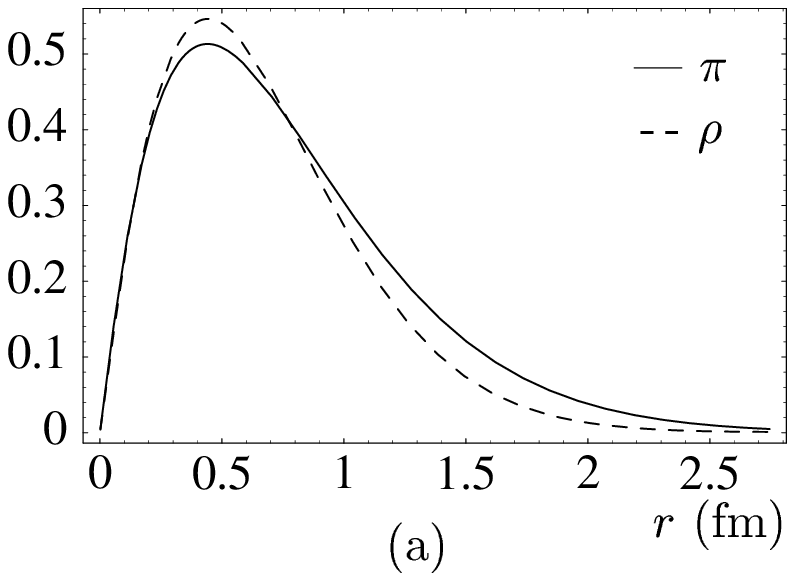}
\qquad\includegraphics[width=7.7cm]{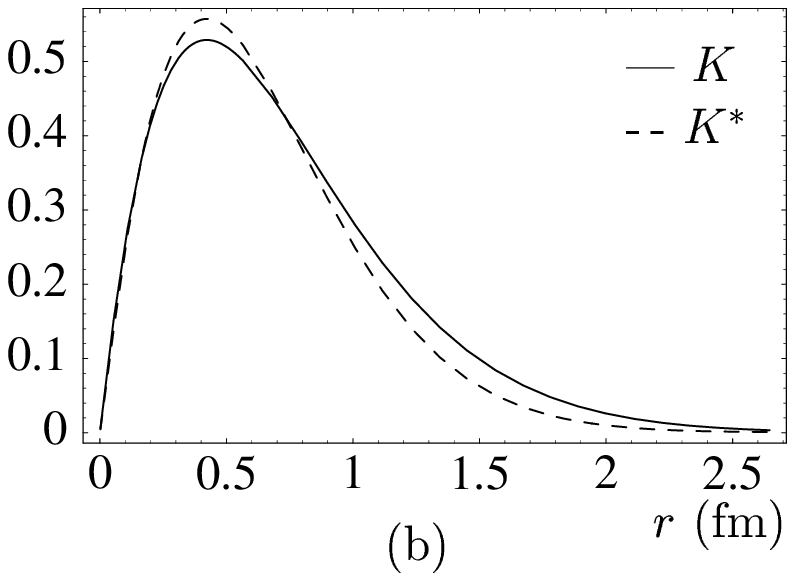}
  \caption{The reduced radial wave functions $R(r)$:  (a) $\pi$
  (solid line) and $\rho$ (dashed line) mesons; (b)~$K$ (solid line)
  and $K^*$ (dashed line)  mesons.}  
  \label{fig:pi}
\end{figure}

The quasipotential equation (\ref{quas}) is solved numerically for the
complete relativistic potential (\ref{eq:v}) which depends on the
meson mass in a complicated highly nonlinear way.  The obtained meson
masses are presented in Table~\ref{tab:mass} in comparison with
experimental data \cite{pdg} and other theoretical results \cite{gi,mr,k}.
This comparison exhibits a reasonably good overall agreement of our
predictions 
with experimental mass values. We consider such agreement to be quite
successful, since in evaluating the meson masses we had at our disposal
only one adjustable parameter  $\Lambda$, which was fixed from fitting
the $\rho$ meson mass. All other parameters are kept the same as in
our previous papers \cite{egf,efg}. The obtained wave functions of
light mesons can be used further for describing their properties
(charge radii, decay constants etc.) and various hadronic matrix elements.   
The wave functions of $\pi$, $\rho$ and $K$, $K^*$ mesons are
shown in Figs.~\ref{fig:pi}(a) and \ref{fig:pi}(b), respectively. The
$\phi$ meson wave function looks similar.

\begin{table}
  \caption{Masses of light $S$-wave mesons (in MeV)}
  \label{tab:mass}
\begin{ruledtabular}
\begin{tabular}{ccccccc}
Meson& State & 
\multicolumn{4}{l}{\underline{\hspace{3.6cm}Theory\hspace{3.6cm}}}
\hspace{-3.6cm}
& Experiment \\
 & $n^{2S+1}L_J$&this work& Ref.\cite{gi}& Ref.\cite{mr}& Ref.\cite{k}&
 PDG \cite{pdg}\\ 
\hline
$\pi$&  $1^1S_1$& 154 &150& 138& 140 &139.57\\
$\rho$ & $1^3S_1$ & 776$^*$ & 770& 742& 785 & 775.8(5)\\
$\pi'$&  $2^1S_1$& 1292 &1300& &1331  &1300(100)\\
$\rho'$ & $2^3S_1$ & 1486&1450& &1420 & 1465(25)\\
$K$ & $1^1S_1$ & 482&470& 497& 506& 493.677(16)\\
$K^*$&$1^3S_1$ & 897&900& 936& 890& 891.66(26)\\
$K'$ & $2^1S_1$ & 1538&1450& &1470&  1460?\\
${K^*}'$&$2^3S_1$ & 1675&1580& &1550& 1717(27)\\
$\phi$& $1^3S_1$& 1038&1020& 1072&990 & 1019.46(2)\\
$\phi'$& $2^3S_1$& 1698&1690& &1472 & 1680(20)
  \end{tabular}
\end{ruledtabular}
\flushleft${}^*$ fitted value
\end{table}

The authors are grateful to D. Antonov, A. Badalian, M. M\"uller-Preussker,
V. Savrin and Yu. Simonov for support and discussions.  Two of us
(R.N.F. and V.O.G.)  were supported in part by the {\it Deutsche
Forschungsgemeinschaft} under contract Eb 139/2-3 and by the {\it Russian
Foundation for Basic Research} under Grant No.05-02-16243. 

%\newpage

\end{document}